\documentclass{aa}  

\usepackage{graphicx}
\usepackage[dvipsnames]{xcolor}
\usepackage{txfonts}
\usepackage[colorlinks=true,
    linkcolor=blue,
    citecolor=blue,
    filecolor=magenta,      
    urlcolor=cyan]{hyperref}
\begin{document} 

   \title{Neural Networks unveiling the properties of gravitational wave background from massive black hole binaries}
   \titlerunning{NN applied to GWB}
   \authorrunning{M. Bonetti et al.}

   \author{
          Matteo Bonetti
          \inst{1,}
          \inst{2,}
          \inst{3}\fnmsep\thanks{matteo.bonetti@unimib.it}
          \and
          Alessia Franchini
          \inst{1,}
          \inst{2}
          \and
          Bruno Giovanni Galuzzi
          \inst{5,4}\fnmsep\thanks{brunogiovanni.galuzzi@ibfm.cnr.it}
          \and
          Alberto Sesana
          \inst{1,}
          \inst{2}
    }

   \institute{
            Dipartimento di Fisica ``G. Occhialini'', Universit\`a degli Studi di Milano-Bicocca, Piazza della Scienza 3, I-20126 Milano, Italy
        \and
            INFN, Sezione di Milano-Bicocca, Piazza della Scienza 3, I-20126 Milano, Italy
        \and
            INAF - Osservatorio Astronomico di Brera, via Brera 20, I-20121 Milano, Italy
        \and
            Dipartimento di Bioscienze e Biotecnologie, Universit\`a degli Studi di Milano-Bicocca, via Roberto Cozzi 53 - 20125 Milano, Italy
        \and
            Istitute of bioimaging and molecular physiology, via Fratelli Cervi, 93 - 20054 Segrate, Italy
    }

   \date{Received xxx / Accepted xxx}

 
\abstract{
Massive black hole binaries (MBHBs) are binary systems formed by black holes with mass exceeding millions of solar masses, expected to form and evolve in the nuclei
of galaxies. The extreme compact nature of
such objects determines a loud and efficient emission of Gravitational Waves (GWs), which can be detected by the Pulsar Timing Array (PTA) experiment in the form of a Gravitational Wave Background (GWB), i.e. a superposition of GW signals coming from different sources.
The modelling of the GWB requires some assumptions on the binary population and the exploration of the whole involved parameter space is prohibitive as it is computationally expensive. We here train a Neural Network (NN) model on a semi-analytical modelling of the GWB generated by an eccentric population of MBHBs that interact with the stellar environment. We then use the NN to predict the characteristics of the GW signal in regions of the parameter space that we did not sample analytically. The developed framework allows us to quickly predict the level, shape and variance of the GWB signals produced in different universe realisations.
}


\maketitle
%
\section{Introduction}

Discovering Massive Black Hole Binaries (MBHBs) is key for our complete understanding of the Universe, in particular how galaxies form and evolve through cosmic time.
Their existence is strongly favored and predicted by all the theoretical models of galaxy mergers \citep[e.g.][]{1980Natur.287..307B}, however, MBHBs still remain elusive with only a number of candidates that are yet to be confirmed.
A very promising and timely way to gain insight into this elusive MBHB population is through the detection of a Gravitational Wave Background (GWB) signal at low (i.e. $10^{-9}-10^{-7}$ Hz) frequencies using millisecond pulsars as macroscopic clocks.

Since pulsars are very precise clocks, tiny deviations from the expected time of arrival of radio pulses on Earth can reveal the passage of GWs \citep{Verbiest2016}. Pulsar Timing Array (PTA) experiments are sensitive to
low frequency GWs emitted by MBHBs with masses above $10^8M_{\odot}$ and whose incoherent superposition is expected to form a stochastic GWB \citep{Lentati2015}.
Very recently, all the PTA collaborations across the globe (i.e. European PTA, Indian PTA, Parkes PTA, North America Nanohertz Observatory for GWs (NANOGrav) and Chinese PTA) have found evidence of a stochastic process with common amplitude and spectral slope across many monitored pulsars with statistical significance between $2\sigma$ and $4\sigma $, depending on the number of pulsars and on the employed analysis technique \citep{2023arXiv230616214A,2023arXiv230616224A,2023arXiv230616225A,2023arXiv230616226A,2023arXiv230616227A,2023arXiv230616228S,nanograv2023,2023ApJ...951L...8A,2023ApJ...951L...9A,2023ApJ...951L..10A,2023ApJ...951L..11A,ppta2023,cptadr1}. 
This discovery clearly opens a completely new window on the Universe, allowing us to deepen our knowledge of different phenomena and probe new astrophysical and cosmological sources.
Several theoretical interpretations of this GW signal hint have been proposed after the discovery \citep{2023arXiv230616227A,2023ApJ...951L..11A}, although the possibly most plausible is a MBHB population origin \citep{2023arXiv230616227A,2023ApJ...952L..37A}. 
Ideally, the signal produced by a population of inspiralling MBHBs would manifest as a stochastic Gaussian process characterized by a power-law Fourier spectrum of
delays-advances to pulse arrival times, with the characteristic inter pulsar correlations identified by \cite{1983ApJ...265L..39H}. 
However, the overall signal might be dominated by a handful of massive, nearby sources that can result in loud enough signals to be individually resolved as Continous GWs \citep[CGW,][]{2009MNRAS.394.2255S,2012PhRvD..85d4034B,2018MNRAS.477..964K}.
Furthermore, the signal might deviate from the isotropy, Gaussianity and stationarity that characterize signals from primordial origin in the Universe \citep{2012ApJ...761...84R}.

The GWB encodes precious information about the elusive MBHB population. In particular, the signal amplitude and spectral shape are deeply connected to the galaxy merger rate and the dynamical properties of the emitting MBHBs \citep{2011MNRAS.411.1467K,2013MNRAS.433L...1S,2014MNRAS.442...56R}. While current models try to describe the GWB in terms of normalisation and spectral shape, the actual MBHB population, being formed by discrete objects, also imprints an intrinsic variance in the GWB.
To model the GWB together with its variance, Monte-Carlo realisations of the entire population (millions of MBHB) are needed \citep{Sesana2008,2010MNRAS.402.2308A,Chen2017}. 
Unfortunately, this makes any wide parameter space exploration computationally prohibitive, allowing limited sampling of GW spectra at discrete points in the multi-dimensional binary population parameter space, while leaving a large fraction of the parameter space unexplored. 
It is then crucial to find a way to ``interpolate'' the GWB spectra across the multi-dimensional space, ultimately extending the investigation through the whole parameter space. An early attempt in this direction was done by \citet{2017PhRvL.118r1102T},  employing a Gaussian Process (GP) emulator to ``interpolate'' over the parameter space represented by two main parameters, initial binary eccentricity $e_0$ and stellar density at influence radius $\rho$.
\cite{2023ApJ...952L..37A} extended the work of \citet{2017PhRvL.118r1102T}, introducing more parameters in the model, but always using a GP-based regression tool. They however neglect the effect of the MBHB eccentricity and focus only on the effects of the environment.

We here propose a new framework where we use a trained Neural Network (NN) to evaluate the GWB in the whole parameter space, beyond the region we use for the training of the NN. 
The advantage of using a NN resides in its deterministic nature and its lower computational cost. NN are also highly flexible and can represent very complex functions with deep architectures. 
We model the MBHB population, taking into account an agnostic parametrisation of the mass function shape and normalisation, the presence of non-negligible eccentricity of the binaries, as well as considering the influence of the stellar hardening on the MBHB evolution.
We train our NN using numerous realizations of the GWB signals in different kinds of ``universes'' in order to efficiently assess their variance.

The paper is organised as follows. In Sec.~\ref{sec:model} we summarise the theoretical framework that we use to model MBHB population, highlighting the novelties in our approach. In Sec.~\ref{sec:NN_setup}, we detail the specifics of the NN algorithm used. In Sec.~\ref{sec:results}, we present the results of our approach, while in Sec.~\ref{sec:conlusions} we discuss the implications of our algorithm, envisage the forthcoming development and usage and finally draw our conclusions.

\section{Theoretical model}
\label{sec:model}

Assuming that the binaries in the Universe are circular and evolve only due to gravitational wave emission, we can write the GWB characteristic amplitude as a function of the number density of mergers  \citep{Phinney2001}
\begin{equation}
    h^2_{\rm c}(f) = \frac{4G}{\pi f^2 c^2}\int_{0}^{\infty} dz \int_{0}^{\infty} d\mathcal{M} \frac{d^2 n}{dzd\mathcal{M}} \frac{1}{(1+z)}\frac{dE_{\rm GW}(\mathcal{M})}{d\ln f_{\rm r}}
    \label{eq:gwb}
\end{equation}
where $\mathcal{M}$ is the chirp mass, the $1/(1+z)$ term takes into account the redshift of gravitons, $d^2n/dzd\mathcal{M}$ is the comoving number density of GW events, usually inferred through numerical simulations or semi-analytical models and $E_{\rm GW}$ is the energy generated by an event. The rest frame frequency is $f_{\rm r}=(1+z)f$, where $f$ is the observed frequency, and $f_{\rm r}=2f_{\rm orb}$, where $f_{\rm orb}$ is the rest frame orbital frequency.

The comoving number density per unit redshift and chirp mass can be written as
\begin{equation}
    \frac{d^2 n}{dzd\mathcal{M}} = \frac{d^3N}{dzd\mathcal{M}d\ln f_{\rm r}}\frac{dz}{dV_{\rm c}}\frac{d\ln f_{\rm r}}{dt_{\rm r}}\frac{dt_{\rm r}}{dz}
    \label{eq:dndzdM}
\end{equation}
where the first term on the r.h.s. indicates the comoving number of binaries emitting in a given logarithmic frequency interval with chirp mass and redshift in the
range $[\mathcal{M},\mathcal{M}+d\mathcal{M}]$ and $[z,z+dz]$ respectively, $dV_{\rm c}$ is the comoving volume shell between $z$ and $z+dz$ and $t_{\rm r}$ is the time measured in the source rest frame.

From cosmology \citep{Hogg1999} we can write \citep[see also][]{Chen2017}
\begin{equation}
    \frac{dz}{dV_{\rm c}}\frac{dt_{\rm r}}{dz} = \frac{1}{(1+z)4\pi c d^2_{\rm M}}
\end{equation}
where $d_{\rm M}$ is the proper-motion distance. For circular binaries, the gravitational radiation is emitted at twice the orbital frequency,
 and the sky- and polarization-averaged strain amplitude is given by \citep{Thorne1987} 
\begin{equation}
    h = \frac{8\pi^{2/3}}{10^{1/2}}\frac{\mathcal{M}^{5/3}}{d_{\rm M}(z)} f_{\rm r}^{2/3}\,.
\end{equation}
The temporal evolution of the emission frequency is expressed as
\begin{equation}
  \frac{dt_{\rm r}}{df_{\rm r}}  = \frac{5c^5\pi^{-8/3}}{96 G^{5/3}}\mathcal{M}^{-5/3} f_{\rm r}^{-11/3}, 
\end{equation}
while the radiated energy per logarithmic frequency interval is given by \citep{Thorne1987}
 \begin{align}
     \frac{dE_{\rm GW}(\mathcal{M})}{d\ln f_{\rm r}} = 
     \frac{\pi^{2/3}G^{2/3}}{3} \mathcal{M}^{5/3} f^{2/3}_{\rm r} = \frac{dt_{\rm r}}{d\ln f_{\rm r}} \pi^2 d^2_{\rm M}(z) f^2_{\rm r} h^2\,.
     \label{eq:engw}
 \end{align}
Substituting Eqs. (\ref{eq:dndzdM}) and (\ref{eq:engw}) in Eq. (\ref{eq:gwb}) we obtain 
\begin{align*}\label{eq:gwb2}
     h^2_{\rm c}(f) = \int_{0}^{\infty} dz \int_{0}^{\infty} d\mathcal{M}  \frac{d^3N}{dzd\mathcal{M}d\ln f_{\rm r}} h^2(f_{\rm r})
\end{align*}
which essentially states that the observed characteristic-squared amplitude of the GWB is given by the integral over all the sources emitting in the frequency bin $d\ln f_{\rm r}$ multiplied by the squared strain of each source \citep{Sesana2008}.
The above expression can be written with a normalization that depends on the details of the MBHB population as\citep{Jenet2006}
\begin{equation}
    h_{\rm c}(f) = h_{\rm 1 yr}\left(\frac{f}{{\rm yr^{-1}}}\right)^{-2/3},
\end{equation}
where $h_{\rm 1 yr}$ depends on the assumed model.

\subsection{Binary population}


We choose a simple and agnostic model for the comoving number density per unit redshift $z$ and binary mass $M$ \citep{Middleton2016}
\begin{equation}
    \frac{d^2 n}{dzdM} = A \left(\frac{M}{10^7 \rm M_\odot}\right)^{-\alpha} e^{-\left(\frac{M}{M_0}\right)^\beta} (1+z)^{\gamma} e^{-z/z_0}\,.
    \label{eq:MBHBdist}
\end{equation}
We assume the parameters $A,\alpha,M_0,\gamma,\beta,z_0$ (i.e. the Universe parameters)\footnote{In the following we fix $\gamma = 1$ and $z_0 = 2$. This choice is motivated by the fact that those quantities weakly affect the mass function and fixing their value allow us to reduce the number of parameters that the NN model needs to account for.} that characterize the number density of the GW sources to vary in a given range, see Tab.~\ref{tab:universes}.
From Eq. (\ref{eq:dndzdM}) we calculate the number of binaries in each frequency bin as
\begin{equation}
    \frac{d^3N}{dzdM_1dqdf_{\rm orb}} =  \frac{d^2 n}{dzdM} (1+z)4\pi c d^2_{\rm M} \frac{dt_{\rm r}}{df_{\rm orb}} 
    \label{eq:population}
\end{equation}
where
\begin{equation}
    \frac{dt_{\rm r}}{df_{\rm orb}}= \left(\frac{df_{\rm orb}}{dt_{\rm r}}\bigg|_{\rm GW}+\frac{df_{\rm orb}}{dt_{\rm r}}\bigg|_{\star}\right)^{-1}
\end{equation}
takes into account the frequency evolution due to GW emission (effective at smaller scales) and stellar hardening (dominating the large scale evolution).

In particular, the contribution due to GW emission for a population of eccentric binaries is given by
\begin{equation}
    \frac{df_{\rm orb}}{dt_{\rm r}}\bigg|_{\rm GW} = \frac{96\, G^{5/3}(2\pi)^{8/3}}{5c^5} \frac{qM_1^{5/3}}{(1+q)^{1/3}} f_{\rm orb}^{11/3} F(e)
    \label{eq:dfdtgw}
\end{equation}
where
\begin{equation}
    F(e) = \frac{1+73/24e^2+37/96e^4}{(1-e^2)^{7/2}}\,.
\end{equation}
The frequency evolution due to the presence of the stellar hardening is given by
\begin{equation}
    \frac{df_{\rm orb}}{dt_{\rm r}}\bigg|_{\star} = \frac{3G^{4/3}M^{1/3}}{(2\pi)^{2/3}}\frac{H\rho}{v_{\rm disp}} f^{1/3}_{\rm orb}
    \label{eq:dfdtstar}
\end{equation}
where $M$ is the binary mass, $H=15$ is a dimensionless constant parametrising the efficiency of energy extraction by stars scattering on a MBHB, $\rho$ is the stellar density and $v_{\rm disp}$ is the velocity dispersion, both quantities calculated at the influence radius of the binary \citep{2015MNRAS.454L..66S}.

\begin{table}
    \centering
    \begin{tabular}{c|c|c}
    \hline
        Quantity & Range & Units \\
        \hline
        \hline
         $ A$ &  $10^{-5}$ -- $10^{-1}$ & Mpc$^{-3}$ Gyr$^{-1}$ \\        
         $\alpha$ & 0 -- 1.5 & - \\
         $\beta$ & 0.5 -- 2 & - \\
         $M_0$ & $10^7$ -- $10^9$ & $M_{\odot}$\\
         $\rho$ & $1$ -- $10^5$ & $M_{\odot}\,{\rm pc}^{-3}$\\
         $e_0$ & 0 -- 0.99 & - \\
         \hline
    \end{tabular}
    \caption{Parameters that characterise a chosen Universe.}
    \label{tab:universes}
\end{table}

Equating Eq.~(\ref{eq:dfdtgw}) and Eq.~(\ref{eq:dfdtstar}), we define the decoupling frequency that marks the transition between the GW and stellar hardening driven regimes:
\begin{equation}
    f_{\rm dec} = \left(\frac{64G^{1/3}(2\pi)^{10/3}}{5c^5} \frac{v_{\rm disp}}{\rho H}\frac{M_1^2 q}{M^{2/3}} F(e_0) \right)^{-3/10}
\end{equation}
where $M_1$ is the mass of the primary, $q=M_2/M_1$ is the mass ratio and $e_0$ is the eccentricity of the binary during the stellar hardening phase that we assume to remain constant until the binary reaches the GW driven regime.
Above the decoupling frequency, the binary circularizes due to GW emission. The orbital frequency and the eccentricity are bound by the following equation
\begin{equation}
    \frac{f_{\rm orb}}{f_{\rm orb,0}} = \left(\frac{1-e_0^2}{1-e^2} \left(\frac{e}{e_0}\right)^{12/19} \left(\frac{1+\frac{121}{304}e^2}{1+\frac{121}{304}e_0^2}\right)^{870/2299}   \right)^{-3/2}\,.
\end{equation}

\subsection{Discrete GWB signal construction}
\label{sec:gwbsignal} 

We have so far assumed that the binary population is described by a continuous differential distribution. However in reality the background is a superposition of discrete contributions from binaries drawn from that continuous distribution. This means that the actual signal fluctuates depending on the specific draw realized in nature, and is this intrinsic variance that we want to properly capture with our approach. To this end, for each value of the universe parameters $(A,\alpha,M,\beta,\rho,e_0)$, we perform 100 realizations of the binary population, sampling the distribution in Eq. (\ref{eq:population}) in a Monte Carlo fashion.

\begin{table}
    \centering
    \begin{tabular}{c|c|c|c}
    \hline
        Quantity & Range & Units & Number of bins \\
        \hline
        \hline
         $z$ &  0 -- 10 & - & 30\\        
         $\log_{10} m_1$ & 7 -- 10.5 & $M_\odot$ & 35 \\
         $\log_{10} q$ & -2 -- 0 & - & 30\\
         $f_{\rm orb}$ & $10^{-11}$ -- $6\times 10^{-8}$ & Hz & 570\\
         $e$ & 0 -- 1 & - & 20 \\
         \hline
    \end{tabular}
    \caption{Grid for binary population sampling. }
    \label{tab:binary_bin}
\end{table}

We characterise the binary population as a function of redshift, primary mass, mass-ratio, orbital frequency and eccentricity. Specifically, we translate Eq.~(\ref{eq:population}) into a numerical distribution function of $(z,M_1,q,f_{\rm orb},e)$ with finite size bins, as detailed in Tab.~\ref{tab:binary_bin}. 
We then sample the discrete number of sources drawing an integer number from a Poisson distribution with mean equal to the non-inter number of binaries in that bin predicted by Eq.~(\ref{eq:population}).

In each multidimensional bin, we draw a random number between the lower and upper limit of $(M_1,q,z,f_{\rm orb},e)$ in order to assign the properties to a binary source. We repeat this procedure $N$ times per bin, where $N$ is the discrete number of binaries in that bin.
In the bins that have $N>50$, we only sample 50 binaries and we then multiply the resulting background by $N/50$.


For each population realisation, the GWB can be computed by summing the GW strain produced by each binary in our population. Circular binaries emit a GW signal at $2f_{\rm orb}$ while eccentric ones also emit at multiple harmonics $f_n = nf_{\rm orb}$, where $n$ is the harmonic number.
The GW strain of each harmonic is given by \citep[see e.g.][]{2010MNRAS.402.2308A}
\begin{equation}
    h_{\rm n}(f_n) = 2\sqrt{\frac{32}{5}}\frac{G^{5/3}}{c^4}\frac{\mathcal{M}^{5/3}}{n d_{\rm M}}\left(2\pi \frac{f_{\rm n}}{n}\right)^{2/3}\sqrt{g(n,e)}
\end{equation}
where the dimensionless function $g(n,e)$ determines the fraction of the GW power that is emitted in each harmonic and reads
\begin{align}
    g(n,e) &= \frac{n^4}{32} \Bigg[\bigg(J_{n-2}(ne)-2eJ_{n-1}(ne)+\frac{2}{n}J_n(ne)\nonumber\\
    &+2eJ_{n+1}(ne)-J_{n+2}(ne)\bigg)^2 \nonumber\\ &+(1-e^2)\Big(J_{n-2}(ne)-2J_n(ne)+J_{n+2}(ne)\Big)^2 \nonumber\\
    &+ \frac{4}{3n^2} J_n^2(ne)\Bigg],
    \label{eq:g_n_e}
\end{align}
with $J_n$ representing the $n$-th order Bessel function of the first kind. 

The relevant frequency band for pulsar timing observations is from $\Delta f = 1/T_{\rm obs}$, where $T_{\rm obs}=30$ yrs is the assumed total observation time, to the Nyquist frequency $1/(2\Delta t)$, where $\Delta t$ is the time between subsequent observations (around a couple of weeks). We assume the observed frequency to vary between $1/T_{\rm obs}$ and $6\times 10^{-8}$ Hz, uniformly spaced by $\Delta f$, yielding $N_f = 56$ values.
The GWB in the frequency bin $[f_j,f_{j+1}]$ is therefore given by the sum of the GW strain multiplied by the number of cycles that each binary makes in the observation time
\begin{equation}
    h_c^2(f_j) = \sum_{i = 1}^N \sum_{n=1}^{\bar{n}} h_{n,i}^2(f_{n,i}) \frac{f_{n,i}}{\Delta f (1+z)} \Theta\left(\frac{f_{n,i}}{1+z}\right)
\end{equation}
where the $1+z$ factor converts from rest frame to observe frequency, $f_{n,i} = n f_{\rm orb,i}$ is the orbital frequency of the $i$-th binary, while $\Theta\left(f_{n,i}/(1+z)\right) =1$ if $f_j \leq f_{n,i}/(1+z) \leq f_{j+1}$ and zero otherwise. Note that the $i$ index runs over the sample of binaries, while the $n$ index runs, for each binary, over the harmonics, with $\bar{n}$ chosen such that a sufficient number of them is included in the computation of the signal. We take $\bar{n} = 3 n_{\rm max}$, where $n_{\rm max}$ is a numerical proxy of the harmonic number at which the maximum GW power is emitted and can be accurately approximated by \citep{2021RNAAS...5..275H} \footnote{For $n > 1000$ we group the harmonics in batches of twenty to speed up the evaluation.}
\begin{equation}
     n_{\rm max}(e) \approx 2 \left(1+\sum_{k=1}^4c_k e^k\right) (1-e^2)^{-3/2}
\end{equation}
where $c_1 = -1.01678, c_2 = 5.57372, c_3 = -4.9271, c_4 = 1.68506$.

Finally, we note that, since we generate 100 realizations of the binary population, we have the same number of GWB signals for each choice of our universe parameters.

\section{Neural Network model}
\label{sec:NN_setup}

A large database of GWB realizations is required for each combination of Universe parameters $(A, \alpha, M, \beta, \rho, e_0)$ to train a Machine Learning (ML) model for regression using a NN.

We created approximately $5 \times 10^5$ GWB signals, corresponding to $N_{\rm model} = 5120$ different universes, with each universe having 100 realizations of the binary population. The parameter values were chosen using Sobol sequences within the ranges reported in Table \ref{tab:universes}. 

We then feed this database, composed of $N_{\rm model} \times N_{f} \times 100$ realizations, into the ML model.
Fixing the Universe parameters, we computed the mean $\mu$ and standard deviation $\sigma$ for each of the $N_{\rm model} \times N_{f}$ distributions. We then created an input-output dataset, where the input consists of the Universe parameters and the frequency, and the outputs are mean and the standard deviations. We employed a Feed-Forward Network, using Keras, with $n_{\rm hidden} + 2$ layers, where $n_{\rm hidden}$ represents the total number of hidden layers, each consisting of $n_{\rm neurons}$ neurons. Each hidden layer is connected to the subsequent one through an activation function $a(x)$. The input layer takes the values of the seven features (six universe parameters + $f$).

We split the dataset into a training set (80\%) and a test set (20\%). The parameters were normalized to have zero mean and unit variance using the values from the training set. The parameters of the NN were estimated using the Adam optimizer with an initial learning rate $l_{\rm r}$, employing a batch size of 128 and 100 epochs. The hyperparameters ($l_{\rm r}$, $a(x)$, $n_{\rm hidden}$, and $n_{\rm neurons}$) were determined using a hyperparameter strategy based on Bayesian optimization, using the Mean Absolute Error (MAE) as the loss function. We employed an early stopping criterion to prevent potential overfitting, halting the training phase when the MAE on a validation set (20\% of the training set) ceases to improve.

It is important to note that the two output parameters (i.e. the mean and the standard deviations) have significantly different orders of magnitude. To account for this discrepancy and ensure effective training, we introduced a weighted loss function. Specifically, we assigned a weight of 10 to the second output parameter and a weight of 1 to the first output parameter in the loss function. This weighting scheme was chosen to ensure that the model optimizes its predictions for both outputs, reaching a balance between their accuracy. We further introduced a weighting function for different frequencies in order to focus the NN on better fitting the lower frequencies where the variance of the GWB signal is generally smaller but most importantly less affected by single source noise.

\section{Results}
\label{sec:results}

We here present the results we obtained first from the MC generation of GWB dataset and then from the training of the NN on our sample.

\subsection{Dataset properties}

\begin{figure}
    \centering
    \includegraphics[width=0.5\textwidth]{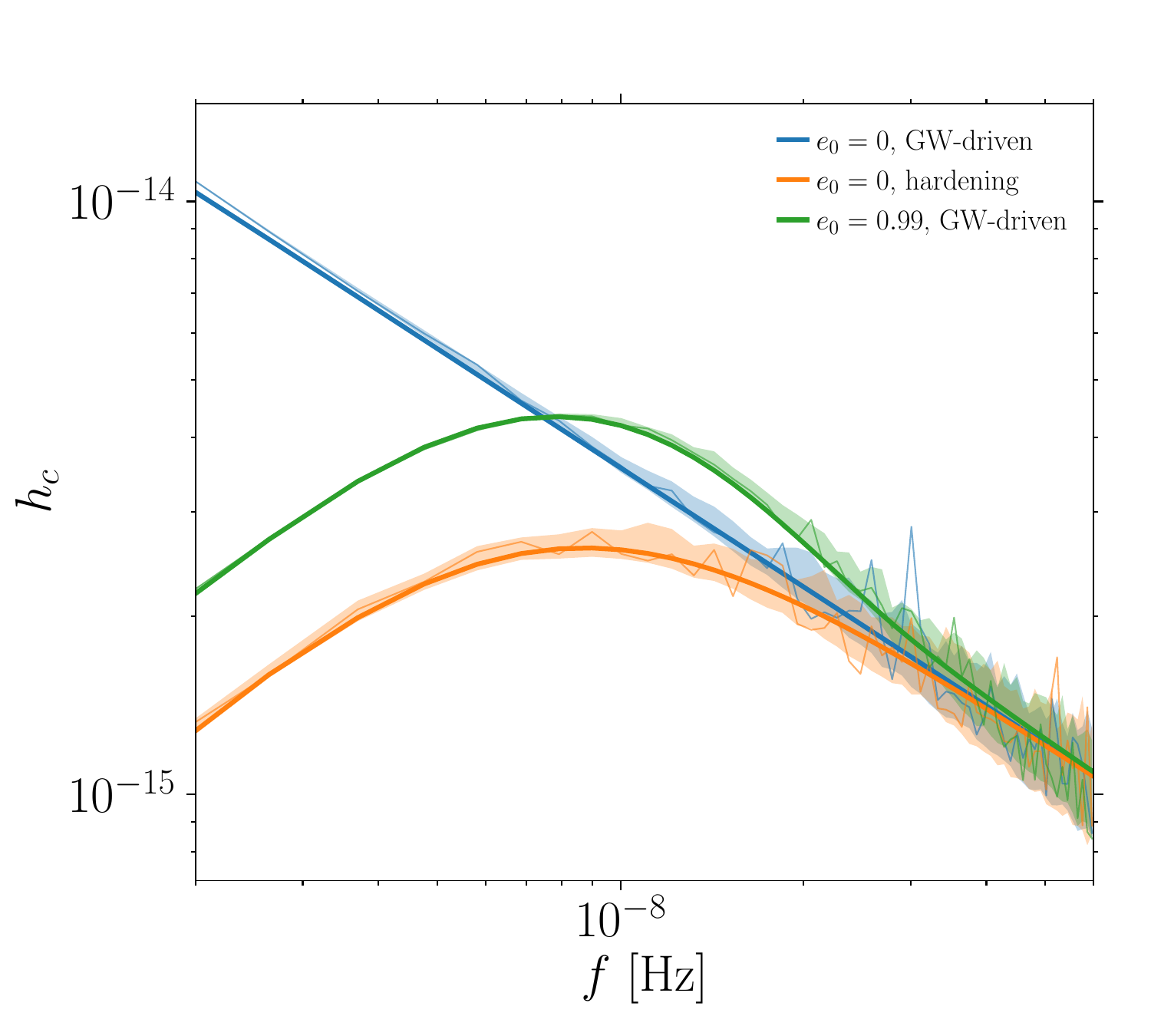}
    \caption{Characteristic strain as a function of frequency for three different universe models ($\log_{10} A = -2, \alpha = 0.3, \beta = 1, \log_{10}M_0 = 8$) featuring: circular GW-driven binaries (blue), circular but stellar+GW-driven binaries ($\rho = 10^4 \rm M_\odot pc^{-3}$, orange), eccentric GW-driven binaries ($e_0 = 0.99 \ @ \ 5\times 10^{-12}$Hz, green). Thick lines denote the theoretical GWB, while shaded areas denote the 10th and 90th percentile of the characteristic strain when different discrete realisation of the population are considered. The thin line with the same color denotes one of such realisation for each model.}
    \label{fig:cfr-gw-hard-ecc}
\end{figure}
\begin{figure}
    \centering
    \includegraphics[width=0.5\textwidth]{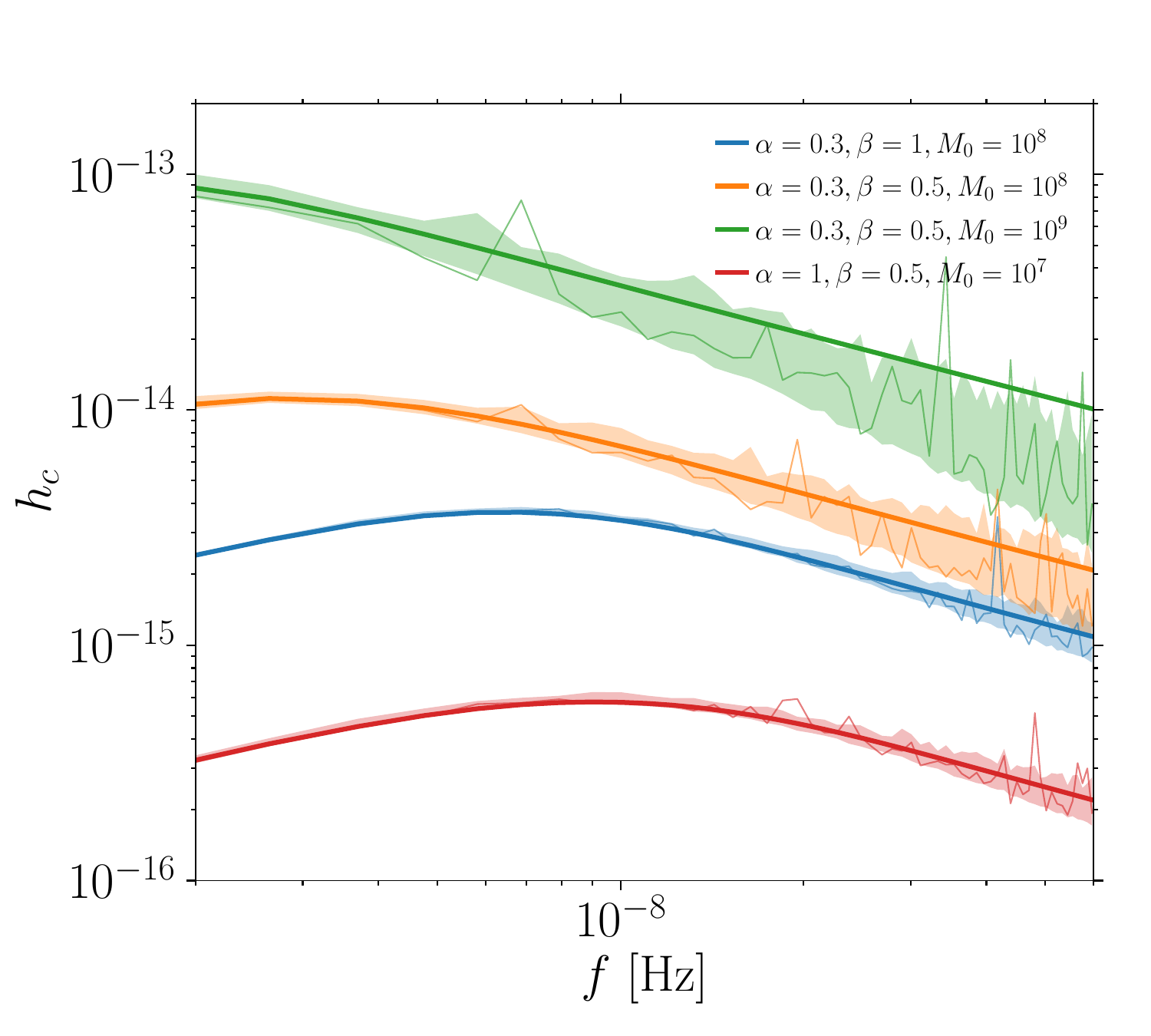}
    \caption{Same as Fig.~\ref{fig:cfr-gw-hard-ecc}, but considering how the characteristic strain of the GWB varies by changing the parameters controlling the shape of the BH mass function (see labels). Higher cut-off masses ($M_0$), shallower power law ($\alpha < 1$) and shallower exponential cutoff ($\beta < 1$) produce more prominent GWB also characterised by a larger variance.}
    \label{fig:alpha-beta-M0}
\end{figure}

Since our dataset is fairly large, it is useful to summarize its main properties and comment on the main features that can be captured by varying different parameters. 

The parameter $A$ acts as a normalisation, shifting up and down the amplitude of the GWB without affecting its shape. Nevertheless, by essentially controlling the number of binaries in a universe, this parameter could in principle be linked to astrophysical relevant parameters. The effective modelling employed in our work can indeed be recast in terms of observable quantities as done in \cite{Chen2017}.

A more substantial role in shaping the GWB is instead played by the parameters $e_0$ and $\rho$. Both parameters affect the GWB spectral shape and are specifically responsible for low frequency turnovers. However, disentangling the effect of stellar hardening from the presence of eccentric binaries is all but straightforward as both effects cause the same similar behaviour, i.e. a turnover in the GWB amplitude at low frequencies. 
The attenuation of the GWB signal is due to the fact that both the presence of a stellar (or gaseous) component and a non-zero initial eccentricity cause the binary to spend less time emitting GWs in a given frequency range. In particular, the interaction of MBHBs with their environment is important at frequencies $f \ll 1 {\rm yr}^{-1}$ where GW emission is still efficient \citep{2011MNRAS.411.1467K}.
Figure \ref{fig:cfr-gw-hard-ecc} shows the signals generated from a population of circular GW-driven binaries (blue line), circular but affected by stellar hardening (orange line) and a population of extremely eccentric GW-driven only binaries (green line). Although both stellar hardening and eccentricity produce a low frequency turnover, there is a clear difference between the orange and green population. While the signal affected by stellar hardening is simply lower than the circular GW-driven one, the GWB generated from eccentric binaries shows both a turnover at low frequencies and a bump which, for the specific model represented, is above $10^{-8}$ Hz \citep[]{2013MNRAS.433L...1S,Chen2017,2017MNRAS.464.3131K}. 
We note here that the turnover at low frequency is significant for very high eccentricities only. Decreasing the eccentricity down to $e_0=0.9$ leads to a much milder effect at low frequencies.
The latter is due to fact that eccentric binaries emit the GW power at multiple harmonics of the orbital frequency, effectively shifting the power from lower to higher frequencies instead of simply depleting the spectrum at lower frequencies by evolving faster as for the stellar hardening. This behaviour of the eccentricity also implies that GWB generated by eccentric MBHB might show a certain degree of correlation among frequency bins, since the same binary can distribute the GW power in several of them.

The most important information for the GWB interpolation across the whole parameter space is how the variance changes as a function of the parameters that characterize the Universe $(A, \alpha,\,\beta,\,M_0,\,\rho,\,e_0)$.
In Figure \ref{fig:cfr-gw-hard-ecc}, the shaded areas denote the intrinsic span of the signal generated by different realisations of the MBHB population. The higher the frequency, the fewer the sources per bin and therefore the stronger is the influence of the granularity of the population on the GWB spectrum, effectively increasing the variance of the signal, possibly featuring strong spikes due to rare but loud binaries. The various effects of MBHB population granularity are more evident from Fig.~\ref{fig:alpha-beta-M0} that shows four different signals generated by changing the Universe parameters that describe the MBHB mass function $\alpha,\,\beta,\,M_0$, which  are responsible for the GWB variance.

\begin{figure*}
    \centering
    \includegraphics[width=0.47\textwidth]{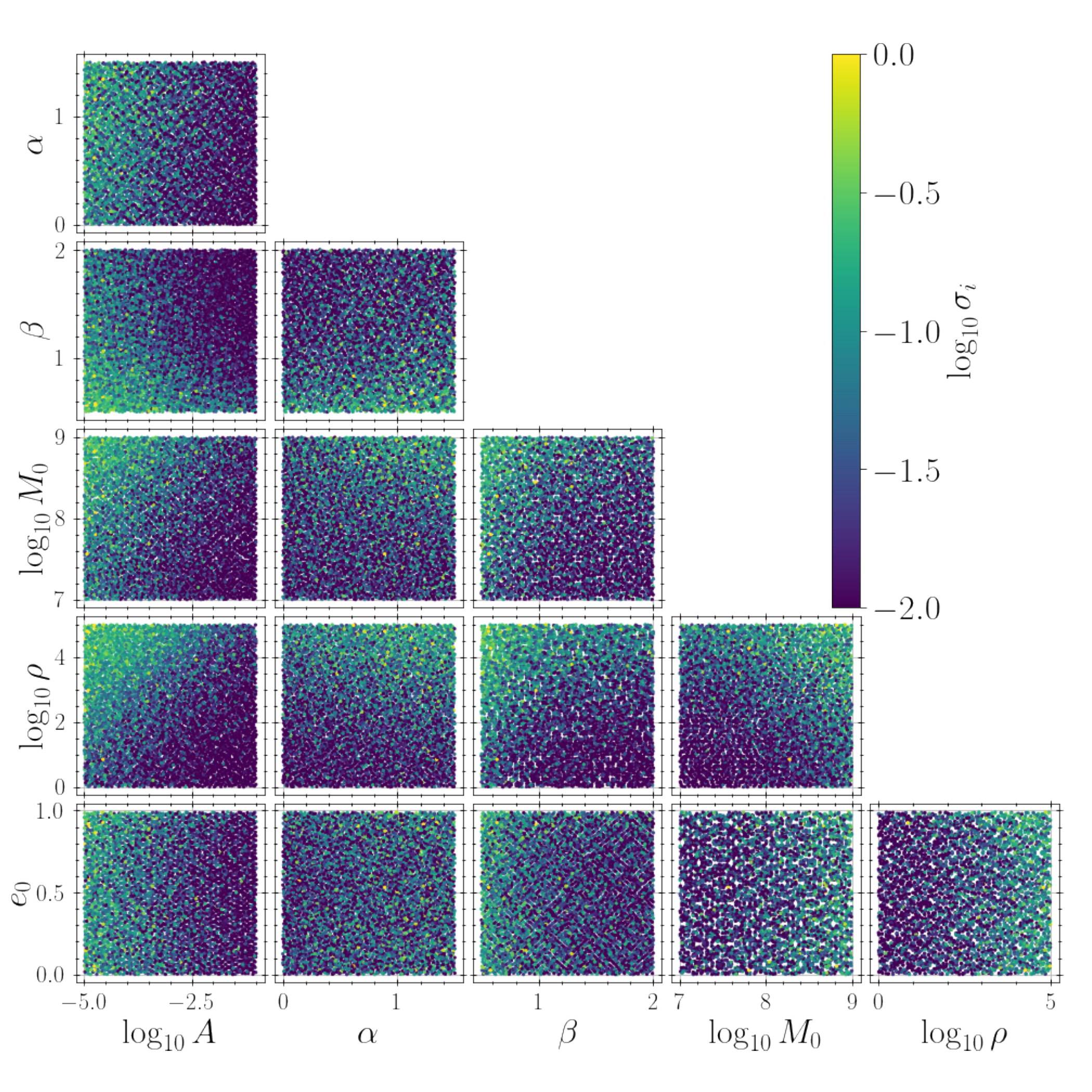}
     \includegraphics[width=0.47\textwidth]{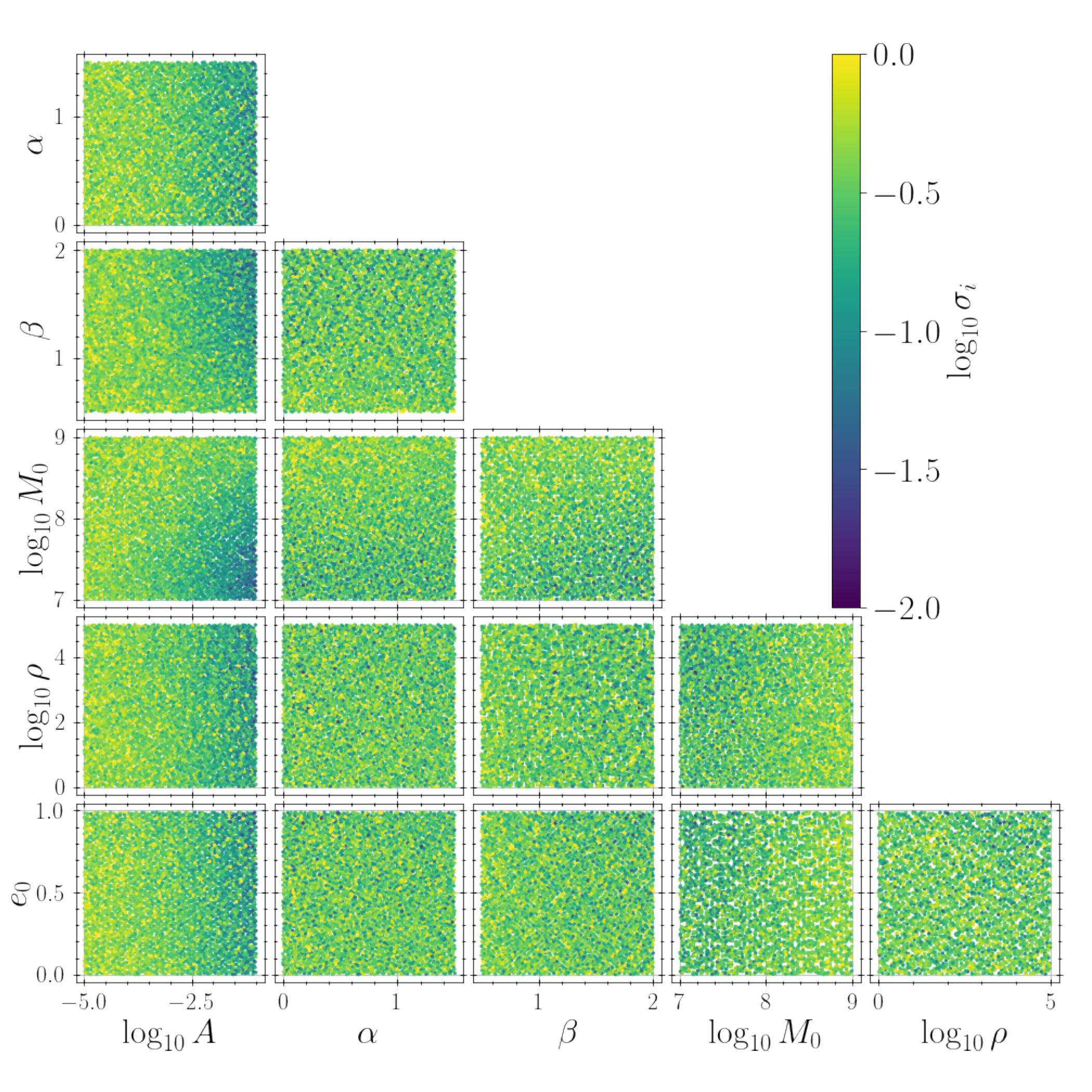}
    \caption{Corner plot showing the dependence of GWB variance (color scale) on the model parameters considering the lowest (left) and highest (right) frequency bin. As can be inferred from the color scale (more evident in the left panel), lower values of $\alpha$ and $\beta$, combined with a high cutoff mass $M_0$ are linked to a larger scatter, in turn caused by the increased probability of having more massive binaries that can produce spikes in the signal. 
    }
    \label{fig:corner_first}
\end{figure*}

A MBHB population characterized by a prominent high mass tail (see green curve of Fig. \ref{fig:alpha-beta-M0}) can drastically affect the GWB spectrum both in terms of normalization and variance. We can see that if the power law decay of the mass distribution is shallower (i.e. $\alpha < 1$) and correspondingly so is the exponential cutoff at high masses ( $\beta<1$ values), the variance between different realizations is larger resulting in a very high scattering in the signal at different frequencies.
The comparison between the green and orange curve shows the effect of increasing the mass cutoff value. An order of magnitude in the cutoff value results in a significantly reduced variance and an order of magnitude weaker signal.
If the MBHB mass function is characterized instead by a steeper exponential decay (blue line, $\beta=1$) but at the same cutoff value (i.e. $10^8M_{\odot}$), the variance is further reduced, the normalization is lower and there is a much more prominent turnover at lower frequencies compared to the shallower exponential decay.
Finally, a steeper power law decay distribution coupled with a smaller mass cutoff value and a shallow exponential decay causes the variance in the signal to drop to a minimum, affecting only the GWB at high frequencies, as shown by the red line. This is consistent with the fact that it is less likely to have very massive MBHBs strongly influencing each GWB signal realization.
However, in this latter case, the amplitude of the signal might be too low to be detectable in the PTA band.

It is also interesting to observe how the GWB variance correlates among the model parameters. However, since all generated GWB have quite different amplitudes we cannot simply compare the different variances (at various frequencies) associated with a specific set of universe parameters $(A, \alpha,\,\beta,\,M_0,\,\rho,\,e_0)$. 
Since however $\rho$ and $e_0$ are the same for all simulations in Figure \ref{fig:alpha-beta-M0}, we can see, from the comparison between the blue and red lines, that the bending occurs at higher frequencies if the signal is strongly dominated by lighter systems (red lines). 
In order to get rid of the GWB normalisation and to focus only on the spread of the signal, at each frequency, we divide the GWB signal of each realization by the mean value (computed over the 100 realisation for a specific set of universe parameters). This allows us to compare the intrinsic spreads of GWBs with different strength levels.
We then collect this information in the corner plots in the left and right panel of Figure \ref{fig:corner_first}, which show with the color scale the ``intrinsic'' variance of the GWB signal in the lowest and highest frequency bin respectively. Darker (lighter) regions correspond to lower (higher) variance.
In general, comparing the two panels of Fig.~\ref{fig:corner_first}, it is clear that the variance is higher at higher frequencies (as expected). 
Moreover, we can also identify regions of the sub-panels showing correlations. Specifically, as already noted above, universes characterised by shallower power-laws (small $\alpha$), shallower exponential cutoff (small $\beta$) and high mass cutoff (large $M_0$) show a clear larger intrinsic variance. Another interesting pattern is shown by the correlation of $A$ and $\rho$, where universes with small $A$ and large $\rho$ seem to show a quite large variance. This is because that combination of parameters reduces quite firmly the number of binaries per frequency bin, increasing therefore the granularity and making that frequency bin more susceptible to influence of rarer but more massive system, possibly producing spikes in the GWB.

\subsection{Model prediction}

\begin{figure*}
    \centering
    \includegraphics[width=\textwidth]{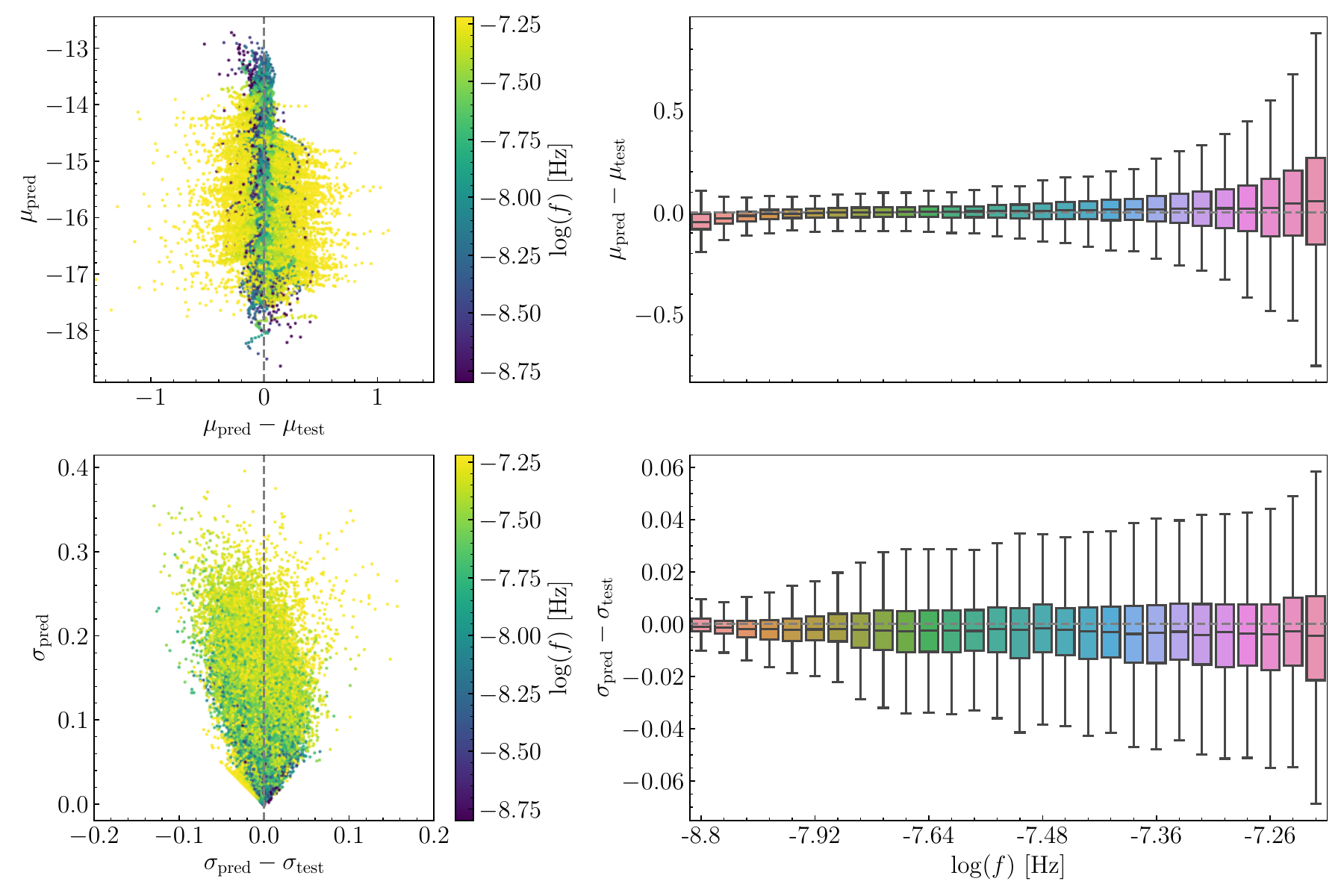}
    \caption{Left panels: absolute error, i.e. the difference between the predicted value and the 
    true value (i.e. that of the test set), as a function of the predicted quantity, either mean $\mu_{\rm pred}$ (top) or standard deviation $\sigma_{\rm pred}$ (bottom). The color scale denotes the frequency bin where the error was computed. Right panels: absolute error on mean (top) and standard deviation (bottom) as a function of frequency. The box plot highlights the median enclosed by the 25th and 75th percentile of the error distribution at each frequency, with the error bars enclosing the full spread of values. Mean and standard deviation are referred to $\log_{10} h_c$. }
    \label{fig:performance}
\end{figure*}

We now focus on the performance of the NN model.
The panels of Figure~\ref{fig:performance} show the
overall accuracy in reproducing the test set mean $\mu$ and standard deviation $\sigma$ for each Universe.  The left panels show how the difference between the predicted value and the true value is distributed in frequency as a function of the predicted quantity itself. The upper left panel represents the mean $\mu$ while the lower left panel represents the standard deviation $\sigma$.
We can clearly see that the distribution of the absolute errors is centred in zero, and that the error on the standard deviation slightly increases for higher predicted values. 

In the same panels, the color scale denotes the frequency at which the comparison is made. It is evident that errors grow with frequency, as expected. This is due to the fact that we assigned, during the training phase, a higher weight to lower frequencies in order to better model the signal in that part of the spectrum. The reason for this choice is twofold: first, the low frequency spectrum is the part that is currently within PTA experiments reach and secondly, the low frequency end of the spectrum is less noisy and therefore easier to model since it is less sensitive to fluctuations generated by loud single sources. This information is better displayed in the right panel of Fig.~\ref{fig:performance}, where the errors on the mean (upper right panel) and standard deviation (lower right panel) are plotted against frequency. At each frequency, the box plot shows the value of the median error limited by the 25th and 75th percentile (color box), with error bars showing the full spread of these values. This provides important information about the distribution of the outliers, as the box size is significantly larger at higher frequencies, where single loud sources dominate the GWB signal. We further note that the errors tend to be smaller at lower frequencies, where we instructed the NN to achieve the best possible performance. 

In Table \ref{tab:performance_tab}, we report some standard accuracy indicators, i.e. the Mean Absolute Error (MAE), the Root Mean Squared Error (RMSE), the square correlation coefficient (R2) and the Spearman correlation (SC), all of them computed on the test set.
All the indicators show the satisfying performance of our NN model. Specifically, the small values of MAE and RMSE imply that the NN model predictions are in agreement with the true values of the test set, while the values of R2 and SC very close to unity for both the mean and standard deviation indicate that the performance of the model is overall very good.

\begin{table}
    \centering
    \begin{tabular}{l|l|l|l|l}
    \hline
         & {\bf MAE} & {\bf RMSE} & {\bf R2} & {\bf SC}  \\ 
         \hline
         $\mu_{\rm pred}$ & 0.0707 & 0.0134 & 0.9791 & 0.9899\\
         $\sigma_{\rm pred}$ & 0.0128 & 0.0003 & 0.9114 & 0.9684 \\
         \hline
    \end{tabular}
    \caption{The Mean Absolute Error (MAE), Root Mean Squared Error (RMSE), square correlation coefficient (R2) and Spearman correlation (SC), computed on the test set.}
    \label{tab:performance_tab}
\end{table}

\begin{figure*}
    \centering
    \includegraphics[width=0.32\textwidth]{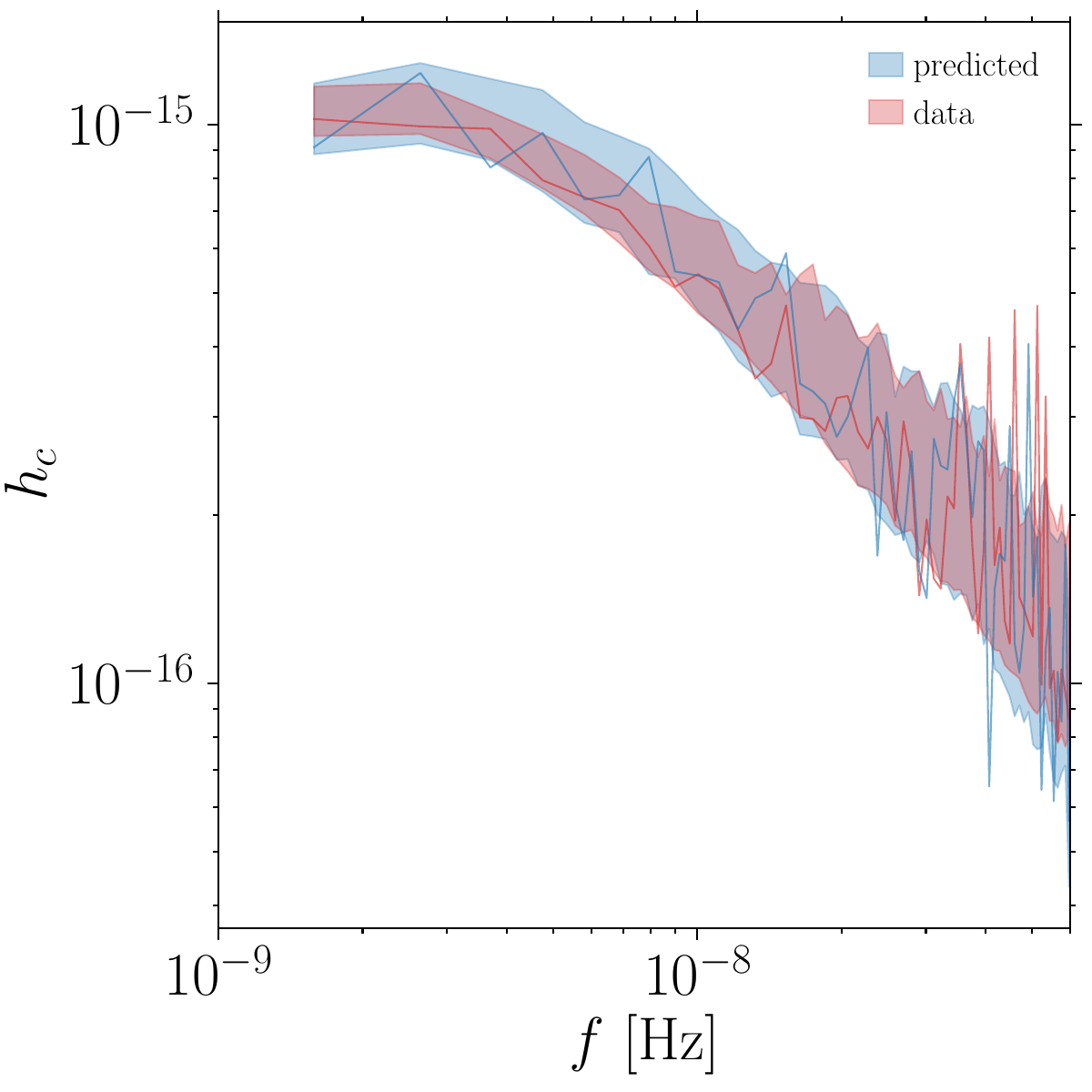}
    \includegraphics[width=0.32\textwidth]{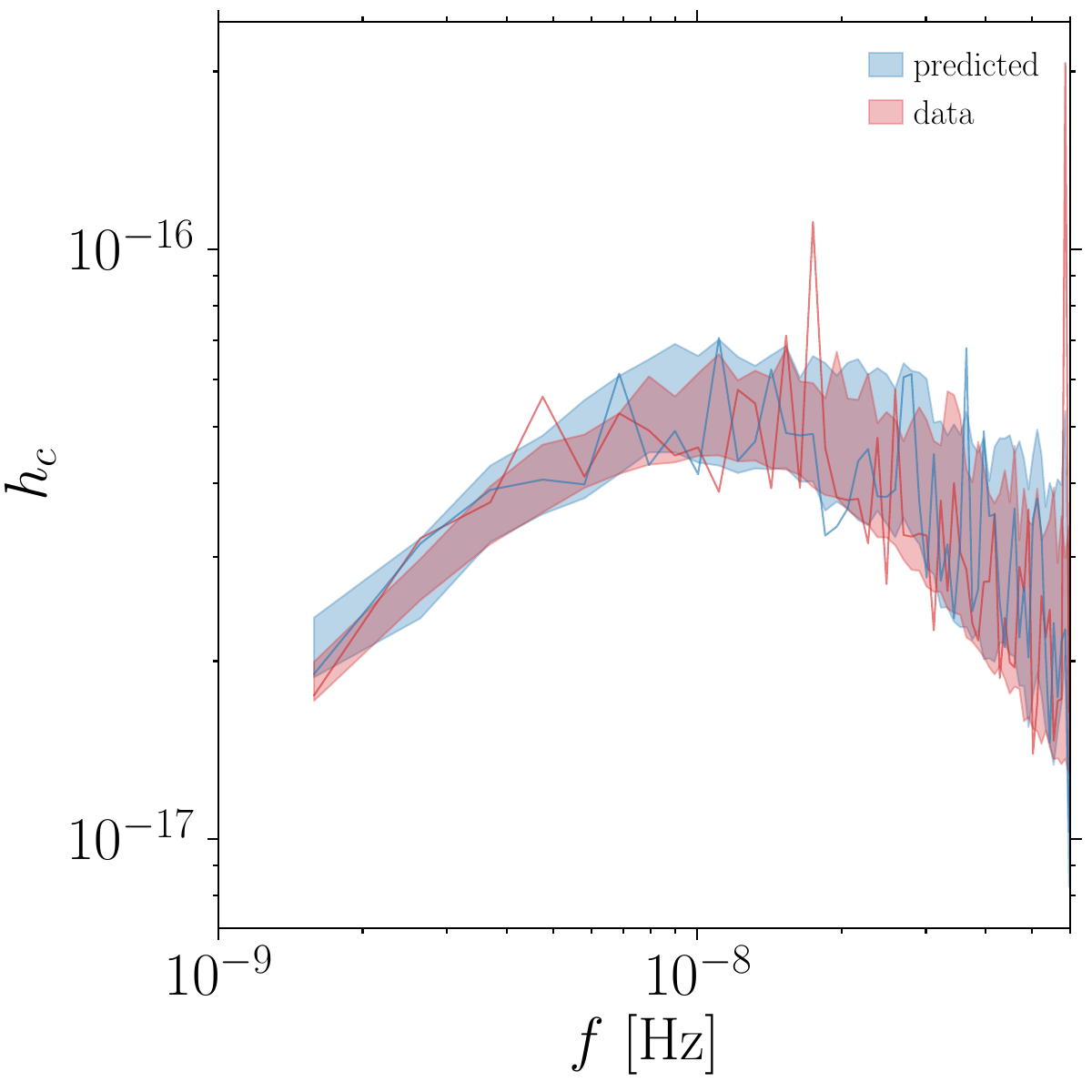}
    \includegraphics[width=0.32\textwidth]{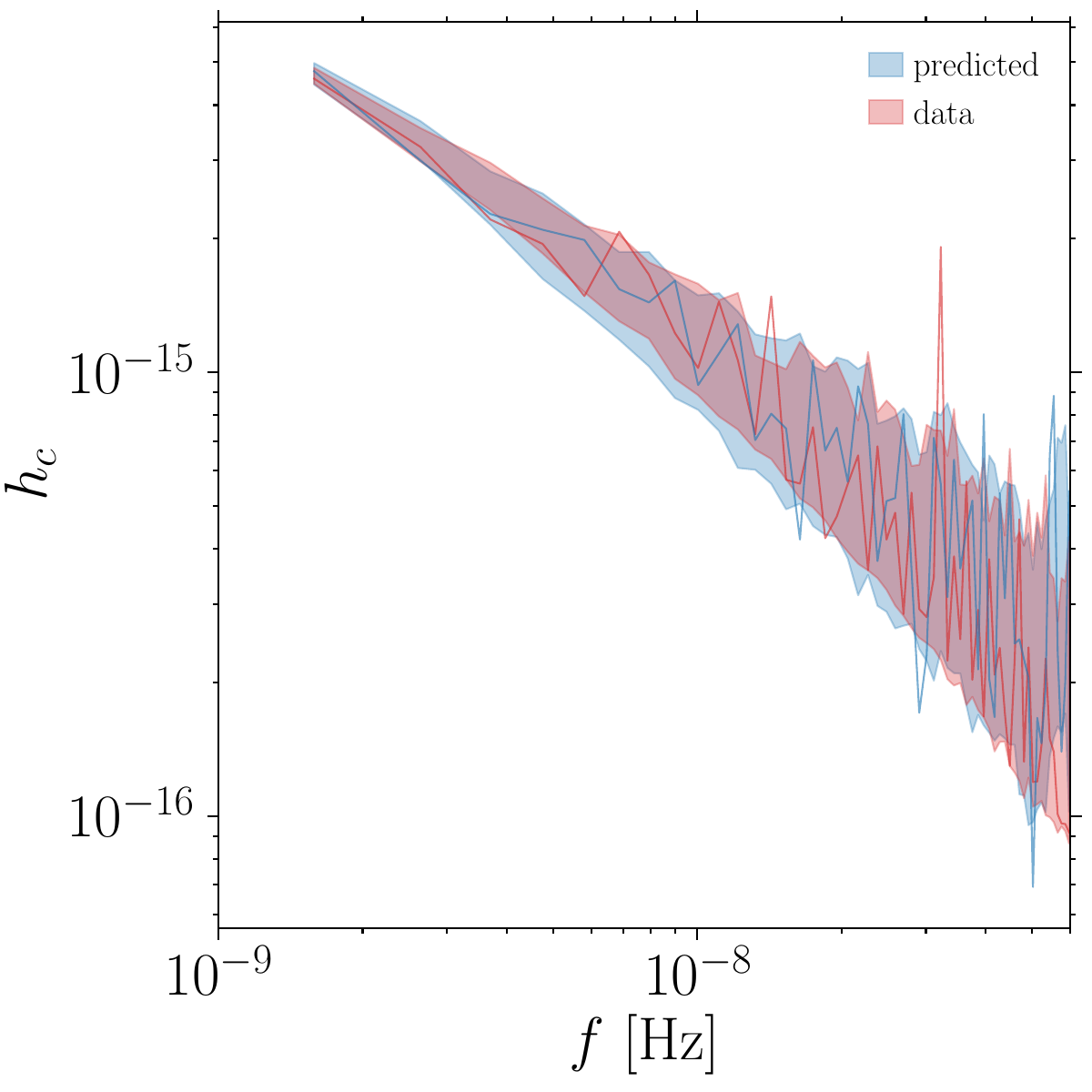}
    \caption{GWB prediction of the NN model compared to three models of the test set of the database. Shaded areas denote the 10th and 90th percentile of the characteristic strain, while the thin lines with the same color represent one GWB realisation for each model. Left panel: universe parameters are $\log_{10}A = -3.39, \alpha = 0.83, \beta = 0.62, \log_{10}M_0 = 8.21, \log_{10}\rho = 2.39, e_0 = 0.19$. Middle panel: universe parameters are $\log_{10}A = -4.19, \alpha = 0.6, \beta = 0.84, \log_{10}M_0 = 7.37, \log_{10}\rho = 3.66, e_0 = 0.17$. Right panel: universe parameters are $\log_{10} A = -4.13, \alpha = 0.28, \beta = 1.42, \log_{10}M_0 = 8.94, \log_{10}\rho = 0.73, e_0 = 0.31$.}
    \label{fig:cfr_model}
\end{figure*}

\begin{figure*}
    \centering
    \includegraphics[width=0.32\textwidth]{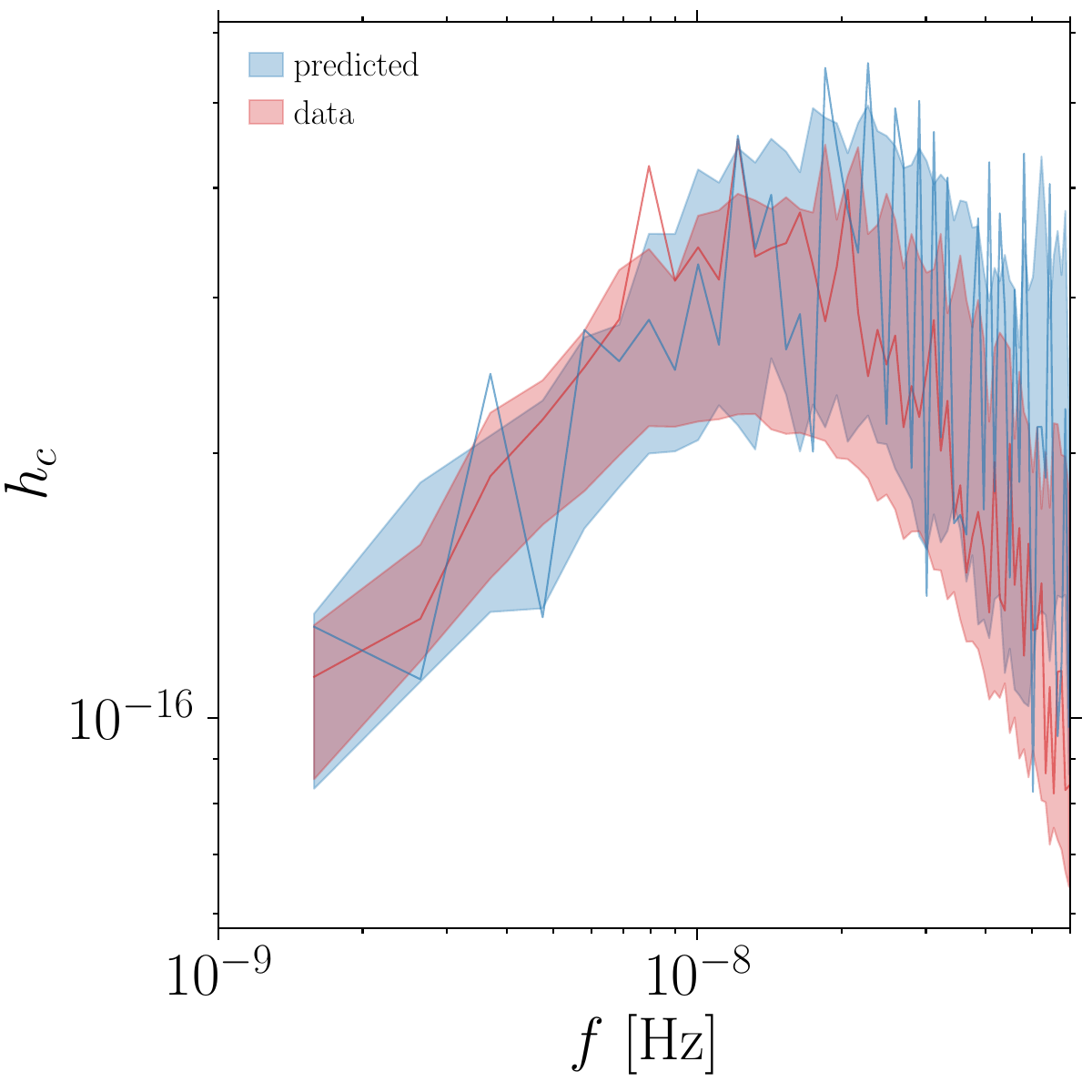}
    \includegraphics[width=0.32\textwidth]{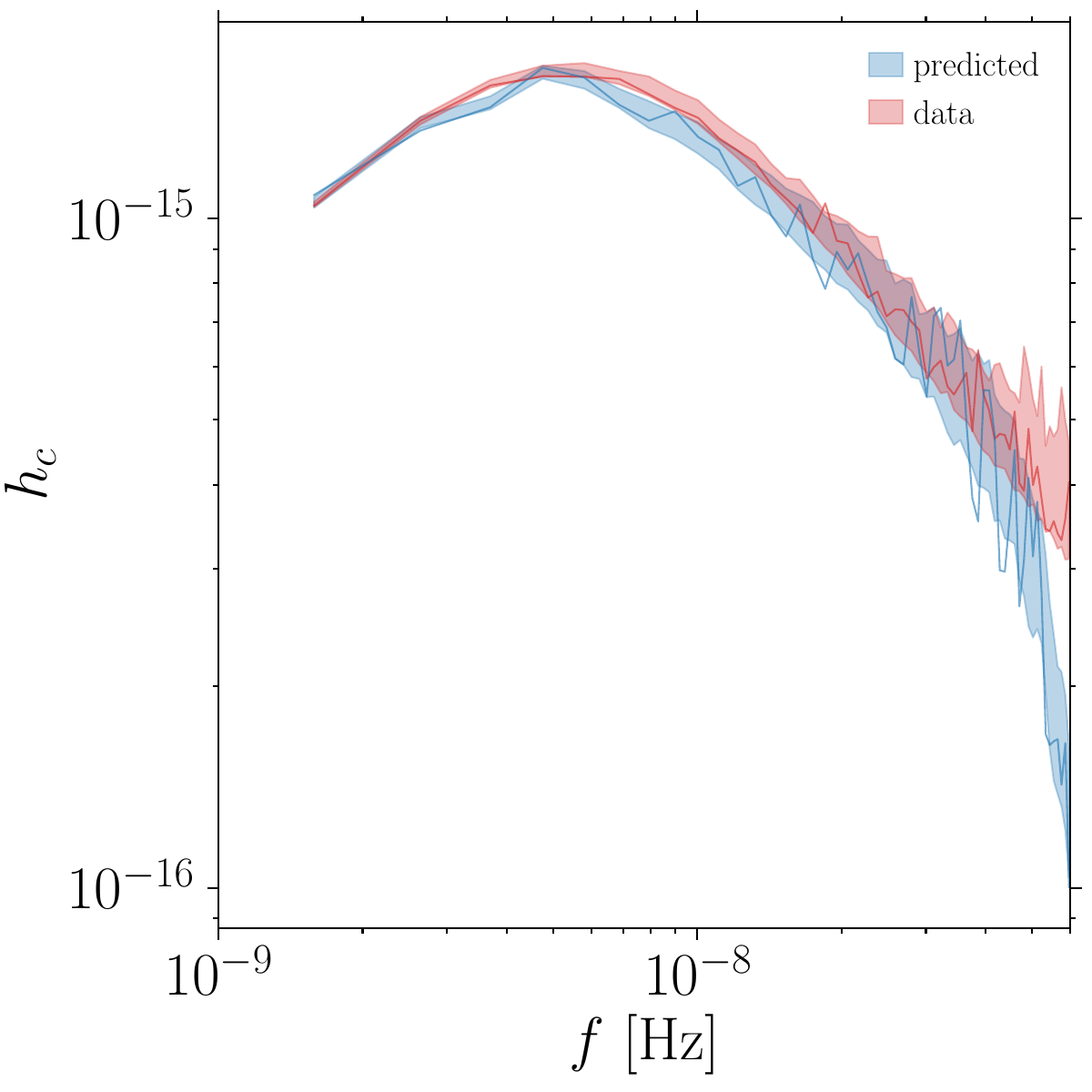}
    \includegraphics[width=0.32\textwidth]{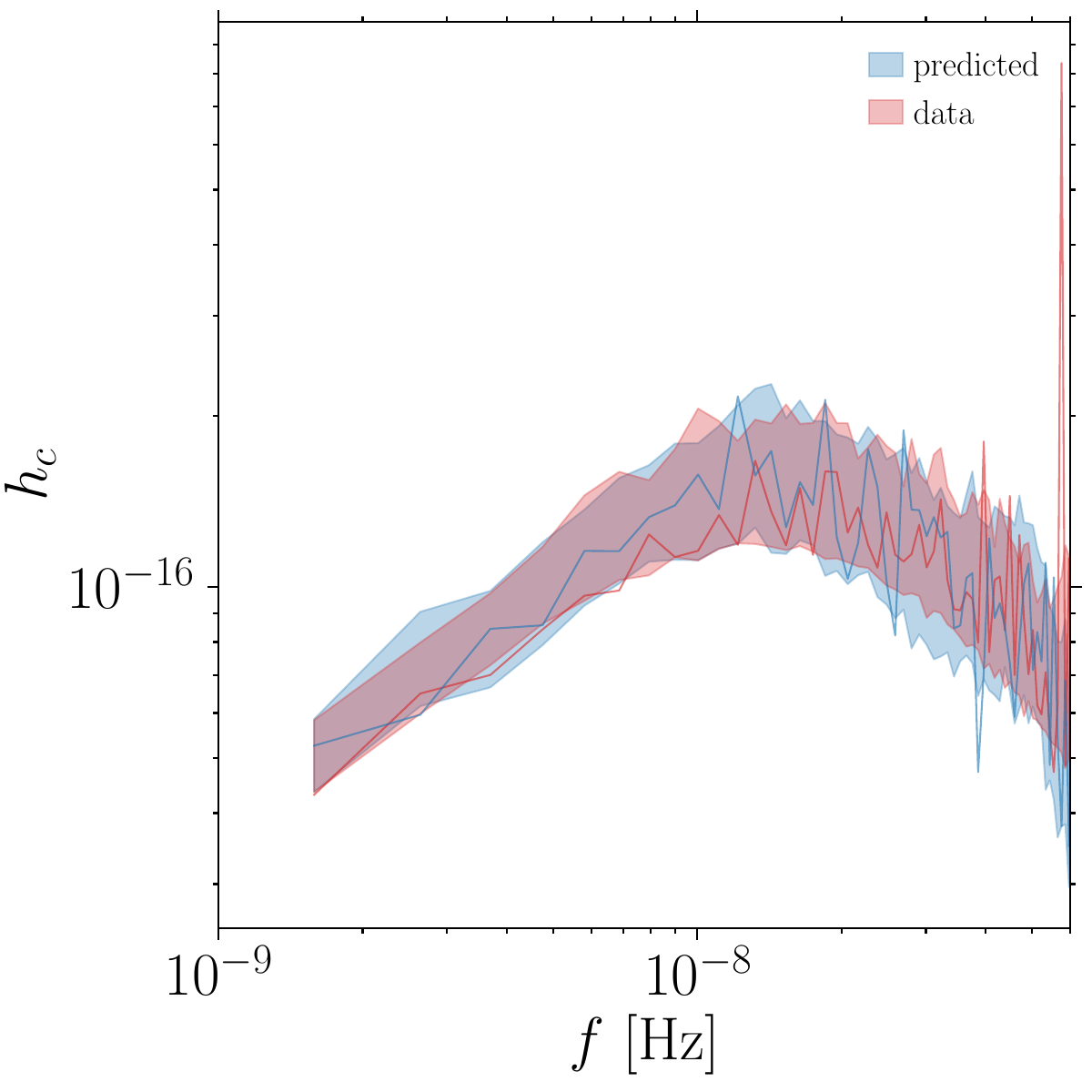}
    \caption{Same as Fig.~\ref{fig:cfr_model}, but for different universe parameters. Left panel: universe parameters are $\log_{10}A = -4.75, \alpha = 0.17, \beta = 1.06, \log_{10}M_0 = 8.70, \log_{10}\rho = 4.34, e_0 = 0.88$. Middle panel: universe parameters are $\log_{10}A = -2.19, \alpha = 1.31, \beta = 1.37, \log_{10}M_0 = 8.29, \log_{10}\rho = 2.00, e_0 = 0.75$. Right panel: universe parameters are $\log_{10} A = -3.90, \alpha = 1.13, \beta = 1.28, \log_{10}M_0 = 8.56, \log_{10}\rho = 3.90, e_0 = 0.8$.}
    \label{fig:cfr_model2}
\end{figure*}

Finally, we represent the predicted GWB for a sample of different Universes in Fig.~\ref{fig:cfr_model} and Fig.~\ref{fig:cfr_model2}. We show the comparison between the GWB constructed following the procedure outlined in Section \ref{sec:gwbsignal} (red) and the signal predicted by our NN framework (blue) for six different choices of the Universe parameters. 
The shaded areas denote the 10th and 90th percentile of the $h_{\rm c}$ distribution at each frequency, while the thin lines represent one realisation of the signal. The NN prediction is in very good agreement with the simulated data, both in terms of the mean, variance and shape of the GWB for all the examples shown in Figs. \ref{fig:cfr_model}-\ref{fig:cfr_model2}. 

We, therefore, conclude that, despite the possible scarcity of performance in a very narrow region of the parameter space, the NN framework here designed predicts very well the shape and strength of the diverse GWB signals.

\section{Discussion and conclusions}
\label{sec:conlusions}

In this work, we built a NN model to efficiently interpolate the stochastic GWB emitted by MBHBs across the wide parameter space that describes their population. We concentrate in particular on the low frequency part of the spectrum as this is currently being surveyed by PTA experiments.

We generated a large dataset of GWB by considering an agnostic modelisation of the underlying MBHB population. GWB are generated from the discrete population of MBHB, significantly improving the simple power-law description of the GWB signal.
We explore $N_{\rm model} = 5120$ different universes (i.e. different parameter configurations), computing 100 realization of the GWB for each of them, to uniformly cover the possible deviations from the power-law prediction due to the eccentricity and hardening of the MBHB population.
This approach allowed us to trace not only the shape and strength but also the variance of the GWB signal, ultimately influenced by the discrete nature of the MBHB population, that can be mainly composed 
by strong signals coming from very massive and/or nearby binaries. 
We note here that we computed the sky-polarization averaged strain, essentially neglecting the effect of the inclination angle of binaries with respect to the line of sight. Taking into account the inclination could potentially introduce more variance as face-on binaries will produce a stronger signal compared to edge-on systems. We however find this effect to be negligible for the background represented by both the green and blue line in Figure \ref{fig:alpha-beta-M0}.

We used the generated dataset to train a NN model to efficiently explore the parameter space, therefore overcoming the bottleneck represented by the expensive MC sampling of realistic MBHB populations for GWB generation.
We found that the NN model performs very well, being able to reproduce the shape, strength and variance of the GWB signals of the test set. We showed that the NN model performs better at lower frequencies, where by design it has been instructed to achieve a better accuracy since current data are now available in that range and the signal is easier to model owing to the lack of strong single source contribution. 

Our trained NN model allows us to efficiently explore the parameter space of our agnostic modelisation. We plan to include the NN model into an end-to-end Bayesian inference pipeline that will be used to produce informed posterior distributions on our model parameters. This will allow us to use the perform astrophysical inference on the signal detected by PTA collaborations, possibly providing constraints on the elusive MBHB populations. 
The {\it ab initio} inclusion of the GWB variance in our modelisation is crucial for the inference of the MBHB parameters. Since we have access to only one Universe (ours), a reliable assessment of the cosmic variance of our MBHB population is of capital importance in order to correctly interpret the nature of the current and future PTA detections. 
We defer the application of our model to PTA data to a forthcoming publication. 

\begin{acknowledgements}
AF, and AS acknowledge financial support provided under the European Union’s H2020 ERC Consolidator Grant ``Binary Massive Black Hole Astrophysics" (B Massive, Grant Agreement: 818691). MB acknowledges support provided by MUR under grant ``PNRR - Missione 4 Istruzione e Ricerca - Componente 2 Dalla Ricerca all'Impresa - Investimento 1.2 Finanziamento di progetti presentati da giovani ricercatori ID:SOE\_0163'' and by University of Milano-Bicocca under grant ``2022-NAZ-0482/B''.
\end{acknowledgements}

%
%

\bibliographystyle{aa} 
\bibliography{biblio}

\end{document}